\documentclass[12pt]{article}
\usepackage{titling}
\usepackage[a4paper]{geometry}
\usepackage[width=.9\textwidth]{caption}
\usepackage{floatrow}
\usepackage{latexsym}
\usepackage{graphicx}
\usepackage{multicol,multirow}
\usepackage{booktabs}
\usepackage{array}

\usepackage{amsmath,amssymb,amsfonts}
\usepackage{mathrsfs}
\usepackage{amsthm}
\usepackage{rotating}
\usepackage{appendix}
\usepackage[square, numbers]{natbib}
\usepackage{ifpdf}
\usepackage[T1]{fontenc}
\usepackage{times}
\usepackage{tabularx}
\usepackage{newtxmath}
\usepackage{textcomp}%
\usepackage{xcolor}%
\usepackage{hyperref}
\usepackage{lipsum}
\usepackage{physics}
\usepackage{authblk}

\newcommand{\M}{\mathcal{M}}
\newcommand{\EE}{\mathbb{E}}
\newcommand{\RR}{\mathbb{R}}
\DeclareMathOperator*{\argmax}{arg\,max}

\title{Statistical constraints on climate model parameters using a scalable cloud-based inference framework}
\date{5 April 2023}

\author[1]{James Carzon \thanks{Corresponding author: jcarzon@andrew.cmu.edu}}
\author[2]{Bruno Abreu}
\author[3,4]{Leighton Regayre}
\author[3]{Kenneth Carslaw}
\author[5,6]{Lucia Deaconu}
\author[5]{Philip Stier}
\author[7,8]{Hamish Gordon}
\author[1,9]{Mikael Kuusela}

\affil[1]{Department of Statistics and Data Science, Carnegie Mellon University, Pittsburgh, Pennsylvania, USA}

\affil[2]{National Center for Supercomputing Applications, University of Illinois Urbana-Champaign, Urbana-Champaign, Illinois,  USA}

\affil[3]{Institute for Climate and Atmospheric Science, School of Earth and Environment, University of Leeds, Leeds, UK}

\affil[4]{Met Office Hadley Centre, Exeter, UK}

\affil[5]{Atmospheric, Oceanic and Planetary Physics Department, University of Oxford, Oxford, UK}

\affil[6]{Faculty of Environmental Science and Engineering, Babes-Bolyai University, Cluj, Romania}

\affil[7]{Department of Chemical Engineering, Carnegie Mellon University, Pittsburgh, Pennsylvania, USA}

\affil[8]{Center for Atmospheric Particle Studies, Carnegie Mellon University, Pittsburgh, Pennsylvania, USA}

\affil[9]{NSF AI Planning Institute for Data-Driven Discovery in Physics, Carnegie Mellon University, Pittsburgh, Pennsylvania, USA}

\begin{document}
\maketitle


\abstract{Atmospheric aerosols influence the Earth's climate, primarily by affecting cloud formation and scattering visible radiation. However, aerosol-related physical processes in climate simulations are highly uncertain. Constraining these processes could help improve model-based climate predictions. We propose a scalable statistical framework for constraining parameters in expensive climate models by comparing model outputs with observations. Using the C3.ai Suite, a cloud computing platform, we use a perturbed parameter ensemble of the UKESM1 climate model to efficiently train a surrogate model. A method for estimating a data-driven model discrepancy term is described. The strict bounds method is applied to quantify parametric uncertainty in a principled way. We demonstrate the scalability of this framework with two weeks' worth of simulated aerosol optical depth data over the South Atlantic and Central African region, written from the model every three hours and matched in time to twice-daily MODIS satellite observations. When constraining the model using real satellite observations, we establish constraints on combinations of two model parameters using much higher time-resolution outputs from the climate model than previous studies. This result suggests that, within the limits imposed by an imperfect climate model, potentially very powerful constraints may be achieved when our framework is scaled to the analysis of more observations and for longer time periods.}

\paragraph{Impact statement}

Atmospheric aerosols influence the amount of solar radiation reflected by Earth, but the magnitude of the effect is highly uncertain, and this is one of the key reasons why climate predictions are highly uncertain. We propose a framework for reducing uncertainty in aerosol effects on radiation by comparing simulations from complex climate models to satellite observations. This framework uses parallel computing and statistical theory to ensure efficient computations and valid inferences.

\section{Introduction}

Atmospheric aerosols affect the formation of clouds and scatter and absorb visible radiation, thereby influencing Earth’s climate. Improved estimates of the change in the total effect of aerosols on the climate over the industrial era -- a highly uncertain quantity termed the aerosol \textit{effective radiative forcing} (ERF) -- have the potential to reduce uncertainty in the sensitivity of the climate to these aerosols. Recent efforts to constrain ERF have involved first reducing uncertainty in the distributions of more basic aerosol-related physical parameters and then studying the effects of these constraints on ERF. This has been notably done by comparing simulated data with observations collected on various global aircraft and ship campaigns as well as from satellites and ground stations \citep{johnson_robust_2020,johnson_importance_2018,regayre_value_2020,regayre_aerosol_2018}. Toward constraining the aerosol radiative forcing, \cite{regayre_identifying_2023} employs simulated outputs from the UKESM1 climate model that are averaged over month-long periods. In contrast, the present work employs three-hourly simulation outputs from the same model, requiring an inferential framework capable of handling this increase in the resolution and quantity of the data.

Establishing constraints on the input parameters of expensive computer models by comparing their outputs with observational data is an area of active research \citep{biegler_large-scale_2010}. It is often imperative that a surrogate model be trained from an ensemble of model input-output pairs and used in place of the simulator to ensure tractable computations, but this step contributes an additional source of uncertainty. If the outputs, or surrogate outputs, do not match observations within some tolerance, then those parameter values are deemed implausible. In \cite{johnson_robust_2020,regayre_value_2020}, the method of \textit{history matching} is used, a technique from oil reservoir engineering which has been adapted to the evaluation of computer models more generally in recent decades \citep{verly_geostatistics_1984, craig_pressure_1997, johansen_statistical_2008, bower_galaxy_2010}. However, the aim of history matching is to constrain parameter spaces and not necessarily to provide well-understood probabilistic guarantees on those constraints.

In contrast with previous work on constraining climate model parameters, our framework draws on a recent surge of interest in simulator-based inference \citep{dalmasso_likelihood-free_2023, cranmer_frontier_2020, schafer_constructing_2009} to produce parameter constraints that provide rigorous statistical guarantees of frequentist coverage. Specifically, our work deals with 
a special case of simulator-based inference where the observations are given by a deterministic simulator and an additive noise model. \cite{patil_objective_2022, stanley_uncertainty_2022} use a \textit{strict bounds} method \citep{stark_inference_1992} to construct efficient confidence sets for the model parameters in closely related inverse problems in remote sensing and high energy physics. Unlike in these works where the forward models of interest are linear and known exactly, the present problem features a forward model (UKESM1) which is nonlinear and estimated using an emulator. We take advantage of the strict bounds method while inverting the emulated forward model numerically and accounting for emulation uncertainty.

Our framework also offers a novel means of accounting for the systematic disagreement, known as the \emph{model discrepancy}, between a simulator and the physical system which it purports to model. A number of approaches to accounting for model discrepancy in computer model calibration or simulator-based inference have been developed \citep{mcneall_impact_2016, higdon_computer_2008, kennedy_bayesian_2001}. We propose a new data-driven procedure for incorporating model discrepancy (and other sources of error that we cannot separately quantify, such as representation errors; \citet{schutgens_spatio-temporal_2017}) into the strict bounds inversion framework. Cloud-based computing resources are leveraged to make each step in the framework computationally scalable.

\subsection{Data sources}\label{sec:data}

Aerosol optical depth (AOD) is a measure for how much aerosol there is in the atmosphere. It is measured by \textit{MODerate-resolution Imaging Spectroradiometer} (MODIS) which is found on board the NASA-launched Terra and Aqua satellites that offer near-global coverage twice daily and provide easily readable open-access data sets. In the flow chart in Figure \ref{fig:data_model}, this data set is the \texttt{SatelliteTimeseries}. For the present application, we focus on MODIS retrievals from the South Atlantic and Central African region over July 1--14, 2017. This domain and period of study is selected for its known biomass burning-related atmospheric aerosol activity.

We compare these data with climate model outputs taken from the UKESM1 model \citep{sellar_ukesm1_2019}. We use simulations that were nudged to the observed meteorology following the method of \cite{telford_technical_2008}, and therefore the simulated weather conditions will be sufficiently realistic that we can examine the ability of the model to represent the observed aerosols. We use the perturbed parameter ensemble (PPE) of 221 atmosphere-only simulations documented by \cite{regayre_identifying_2023}, which have a configuration that closely matches the atmosphere component of the UKESM1 model used in the CMIP6 experiments \citep{sellar_ukesm1_2019}. For each ensemble member, let $u$ be a vector of parameter inputs which determine climatic aerosol-related processes of practical importance, and let $x$ be a vector of control variables which define the specific output of the climate model, denoted $\eta(x, u)$, representing the climate observable $\zeta(x)$. In our setting, $x$ denotes a latitude-longitude-time triple in the climate model output's spatiotemporal grid, denoted $\M_{\text{sim}}$ and called the \texttt{SimulatedGrid} in Figure~\ref{fig:data_model}. The parameters $u$ are listed in Table~\ref{tab:parameters}. The ensemble is a set of simulations, in notation
\begin{equation*}
    D_{\text{train}}=\{(x, u^j, \eta(x, u^j)) : x\in \M_{\text{sim}}, j=1,2,\ldots,221\}.
\end{equation*}
For reference, this and some later notation used throughout this paper is summarized in Table~\ref{tab:notation}.

\begin{table}[]
    \centering
    \begin{tabular}{p{0.2\textwidth}>{\raggedright}p{0.4\textwidth}>{\centering}p{0.1\textwidth}>{\centering\arraybackslash}p{0.1\textwidth}}
        \toprule
        && \multicolumn{2}{c}{Range} \\
        \cmidrule(r){3-4}
        Parameter name    & Physical name & Min. & Max. \\
        \midrule
        sea\_spray & Sea spray emission flux SF & 0.25 & 4.00 \\\hline
        
        bl\_nuc & Boundary layer nucleation rate SF & 0.1 & 10 \\\hline
        
        ait\_width & Aitken mode width (nm) & 1.2 & 1.8 \\\hline
        
        cloud\_ph & Cloud droplet pH & 4.6 & 7 \\\hline
        
        prim\_so4\_diam & Median diameter of primary ultrafine anthropogenic sulfate particles (nm) & 3 & 100 \\\hline
        
        anth\_so2\_r & Anthropogenic SO$_2$ emissions flux SF outside of Europe, Asia and North America & 0.6 & 1.5 \\\hline
        
        bvoc\_soa & Biogenic secondary organic aerosol from volatile organic compounds SF & 0.32 & 3.68 \\\hline
        
        dms & Dimethyl sulfide emission flux SF & 0.33 & 3.0 \\\hline
        
        dry\_dep\_ait & Aitken mode aerosol dry deposition velocity SF & 0.5 & 2.0 \\\hline
        
        dry\_dep\_acc & Accumulation mode aerosol dry deposition velocity SF & 0.1 & 10.0 \\\hline
        
        dry\_dep\_so2 & SO$_2$ dry deposition velocity SF & 0.2 & 5.0 \\\hline
        
        bc\_ri & Imaginary part of the black carbon refractive index & 0.2 & 0.8 \\\hline
        
        a\_ent\_1\_rp & Cloud top entrainment rate SF & 0 & 0.5 \\\hline
        
        autoconv\_exp\_nd & Exponent of $N_d$ in power law for initiating autoconversion of cloud drops to rain drops & -3 & -1 \\\hline
        
        dbsdtbs\_turb\_0 & Cloud erosion rate ($s^{-1}$) & 0 & 0.001 \\\hline
        
        bparam & Coefficient of the spectral shape parameter (beta) for effective radius & -0.15 & -0.13 \\\hline
        
        carb\_bb\_diam & Carbonaceous biomass burning primary particle median diameter (nm) & 90 & 300 \\
        \bottomrule
    \end{tabular}
    \caption{Seventeen of the 37 UKESM1 parameters \citep{regayre_identifying_2023} used to build the surrogate model, selected based on relevance for predicting AOD. $N_d$ is the cloud droplet number concentration. SF is short for \textit{scale factor}.}
    \label{tab:parameters}
\end{table}

\begin{table}[]
    \centering
    \begin{tabular}{ll}
        \toprule
        Notation    & Meaning \\
        \midrule
        $u$ & Vector of parameter values from $\RR^p$ \\\hline
        $x$ & Location in space and time of a measurement \\\hline
        $\M_{\text{sim}}$ & \texttt{SimulatedGrid} (as in Figure~\ref{fig:data_model})  \\\hline
        $\M_{\text{sat}}$ & \texttt{SatelliteGrid} \\\hline
        $\M$ & \texttt{MatchedGrid} \\\hline
        $\M^*$ & \texttt{MatchedGrid}, excluding outliers \\\hline
        $z$ & AOD observations \\\hline
        $\zeta$ & True climate system \\\hline
        $\eta$ & Climate model \\\hline
        $\Tilde{\eta}$ & Emulator for the climate model \\\hline
        $D_{\text{train}}$ & Triplets $(u, x, \eta(x,u))$ \\\hline
        $D_{\text{test}}$ & Tuples $(u, x, \EE\left[\Tilde{\eta}_x(u)\vert D_{\text{train}}\right], \text{Var}[\Tilde{\eta}_{x}(u) \,\vert\, D_{\text{train}} ])$ \\
        \bottomrule
    \end{tabular}
    \caption{A notational reference table.}
    \label{tab:notation}
\end{table}

\subsection{Problem setup}

In order to emulate the model output for unobserved parameter values, we assume as in \cite{johnson_robust_2020} that at $x\in\M_{\text{sim}}$, the model output is a realization of a Gaussian process,
\begin{align}
    \Tilde{\eta}_x(u) &\sim \mathcal{GP}[m_x(\cdot), k_x(\cdot, \cdot)]. && (u\in\RR^p) \label{eq:GP}
\end{align}
For each $x$, we train the surrogate model $\Tilde{\eta}_x$ as described in Section \ref{sec:emulate}.

Let $z(x)$ denote the AOD observations and $u^*$ the true parameter value. Assuming that the emulated climate model is unbiased, these observations and parameters are related according to the equations
\begin{align*}
    z(x) = \zeta(x) + \epsilon_{\text{meas},x}
        = \EE\left[\Tilde{\eta}_x(u^*)\vert D_{\text{train}}\right] + \epsilon_{\text{emu},x}(u^*) + \epsilon_{\text{meas},x} + \epsilon_{\text{other},x}.
\end{align*}
The different sources of variance in the measurements -- namely, the measurement uncertainty ($\epsilon_{\text{meas}, x}$), the emulation uncertainty ($\epsilon_{\text{emu}, x}$), and any other sources ($\epsilon_{\text{other}, x}$) which are not analyzed uniquely, including uncertainty due to model discrepancy, including erroneously simulated meteorology, error in representativeness of measurements, etc. -- are assumed to be mean zero and independent across $x$. This is a simplifying assumption in that, in reality, these terms might be correlated across $x$. By further assuming that $\epsilon_{\text{meas}, x}$ and $\epsilon_{\text{other}, x}$ are Gaussian, the observations $z(x)$ are jointly normally distributed across the $x$ on the spatiotemporal grid $\M_{\text{sat}}$ (\texttt{SatelliteGrid}) on which the MODIS retrivals are gridded. In particular,
\begin{equation}\label{eq:data_model}
    z = (z(x))_{x\in\M_{\text{sat}}} \sim N\left(\EE\left[(\Tilde{\eta}_x(u^*))_{x\in\M_{\text{sat}}}\vert D_{\text{train}}\right], \Sigma_{\text{meas}} + \Sigma_{\text{emu}} + \delta^2 I_{\lvert\M_{\text{sat}}\rvert}\right),
\end{equation}
where $\Sigma_{\text{meas}}$ and $\Sigma_{\text{emu}}$ are covariance matrices such that entry $\Sigma_{\text{meas}, i, j} = \text{Var}(\epsilon_{\text{meas}, x^i})$ is the measurement uncertainty at location $x^i$ when $i=j$, otherwise zero (i.e., it is a diagonal matrix and the measurement errors are assumed to be uncorrelated between locations); $\Sigma_{\text{emu}, i, j}=\text{Var}(\epsilon_{\text{emu}, x^i}(u^*))$ is the emulation uncertainty when $i=j$, otherwise zero; and $\delta^2 = \text{Var}(\epsilon_{\text{other, x}})$ is a homoscedastic variance term standing in for all other unaccounted-for errors. The matrix $I_{\lvert\M_{\text{sat}}\rvert}$ is an identity matrix with number of rows equal to the size of the set $\M_{\text{sat}}$. The disagreement between grids $\M_{\text{sim}}$ and $\M_{\text{sat}}$ is addressed in the following section. The $\Sigma_{\text{emu}, i, i}$ are modeled by the surrogate model (see Section \ref{sec:emulate}). The $\Sigma_{\text{meas}, i, i}$ are the published MODIS uncertainties, which may not account for all possible problems with the retrievals, but these unaccounted-for uncertainties will be captured by $\delta^2$, which is estimated from the observations (see Section \ref{sec:test}).

\section{Inference framework}

\begin{figure}
    \centering
    \includegraphics[width=0.9\textwidth,keepaspectratio]{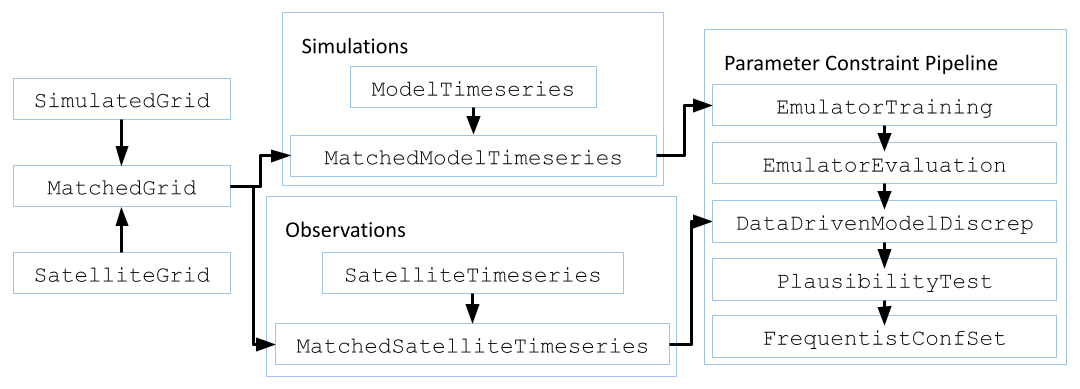}
    \caption{The flow chart for our pipeline for building frequentist confidence sets on climate model parameters. After matching both the satellite observation and model output grids, five steps of processing follow. The \texttt{EmulatorTraining} and \texttt{EmulatorEvaluation} pipes provide scalability to the framework by leveraging parallel computing in these most expensive steps. The \texttt{DataDrivenModelDiscrep}, \texttt{PlausibilityTest}, and \texttt{FrequentistConfSet} pipes implement the strict bounds based method to ensure principled uncertainty quantification}
    \label{fig:data_model}
\end{figure}

We use the C3.AI Suite, a cloud computing platform for data analytics workflows deployed to Microsoft Azure infrastructure \cite{c3_ai_2022}. The platform combines databases, open-source packages, and proprietary machine-learning workflows optimized for working with large-scale, data-intensive applications. We built new data structures and methods for processing NETCDF4 files containing high-dimensional time-series datasets. We also developed a scalable inference pipeline for training and predicting through several thousands of Gaussian process models using asynchronous processing such as parallel batch and map-reduce jobs. This pipeline is summarized in Figure \ref{fig:data_model}, and complements other recently published workflows for similar tasks~\citep[e.g.,][]{watson-parris_model_2021}.

The grid of satellite measurements is finer than the grid of simulations. The raw MODIS retrievals are pre-processed algorithmically by the MODIS team before they are provided on a $1^\circ\times1^\circ$ spatial grid in the Level-3 Global Gridded Atmosphere Product \citep{hubank_modis_2020}. However, the model outputs are on a $1.35^\circ\times 1.875^\circ$ grid. To reconcile these differences, we match simulated grid cells to their nearest neighbor on the Level-3 MODIS product grid in space and time. Differences are computed on the resulting $1.35^\circ\times 1.875^\circ$ resolution \texttt{MatchedGrid}, denoted $\M$.

\subsection{Emulate: Inside the \texttt{EmulatorTraining} pipe}\label{sec:emulate}

As indicated in Eq.~\eqref{eq:GP}, we assume the climate model is a realization of a Gaussian process where for each $x\in\M$ the mean and anisotropic exponential covariance functions are
\begin{align*}
    m_x(u) = \beta_{0, x}, \qquad
    k_x(u, u') = \beta_{1, x}^2\exp\left(-\sqrt{\sum_{i=1}^p \frac{(u_i-u_i')^2}{\ell^2_{i, x}}}\right).
\end{align*}
We make this choice for $k_x$ for the reason that a relatively rough process appears reasonable in this problem. We place an anisotropy assumption on the model by fitting a different length scale parameter $\ell_{i, x}$ for each model parameter $u_i$. By fitting $\ell_{i, x}$ separately for each $x$, we let the parameters' effects on the model output vary geographically.

\begin{figure}
    \centering
    \includegraphics[width=\textwidth,keepaspectratio]{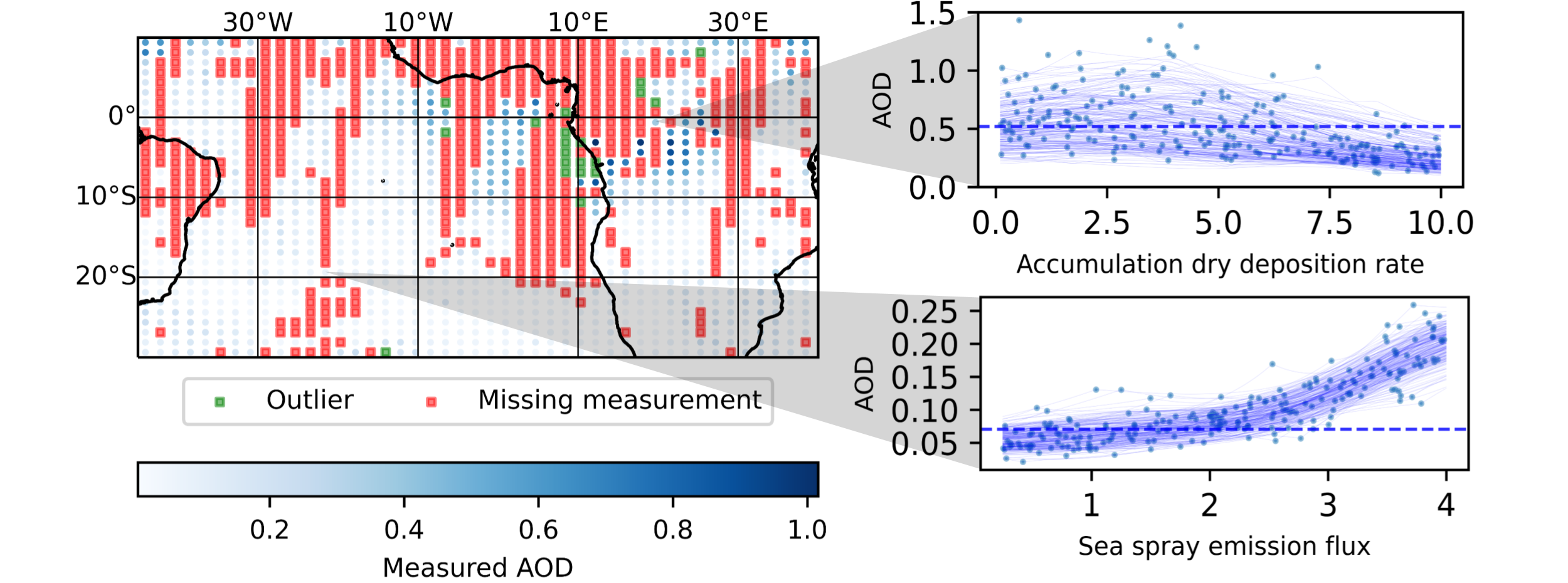}
    \caption{Sample curves of the emulated response $\EE[\Tilde{\eta}_x(u)\,\vert\, D_{\text{train}}]$ averaged over two MODIS observing times on July 1, 2017 for two locations $x$. (Left) Red gridpoints are missing MODIS AOD retrievals. Green gridpoints are ruled out as outliers per Section \ref{sec:predict}. (Top right) The scattered points are from $D_{\text{train}}$, and the 221 curves are slices of the trained emulator response surface where all of the parameters are fixed to their training values from $D_\text{train}$ except the parameter labeling the $x$ axis of each subplot, which is varied within its range given in Table \ref{tab:parameters}. Near $(0^\circ, 20^\circ)$, AOD decreases as the accumulation dry deposition rate increases. The average MODIS measurement is given by the dashed line. (Bottom right) At $(-20^\circ, -20^\circ)$, emulated AOD responds positively to the sea spray emission flux}
    \label{fig:marg_response}
\end{figure}

We train the emulating Gaussian processes by estimating their parameters on $D_{\text{train}}$ using maximum likelihood \citep{rasmussen_gaussian_2006} with the scikit-learn Python library and L-BFGS-B optimization algorithm, and thus we obtain a collection of models $\Tilde{\eta}_{x}$, $x\in\M$. Figure \ref{fig:marg_response} illustrates the quality of these emulators by verifying that the emulated AOD varies with respect to active parameters exactly as the training data set would suggest. In terms of scalability, an ordinary Gaussian process model on the entire \texttt{ModelTimeseries} would compute in $O\!\left((\lvert\M\rvert n)^3\right)$ time, $n$ being the number of members in the PPE, whereas the \texttt{EmulatorTraining} pipe trains in $O\!\left(\lvert\M\rvert n^3\right)$ time. This routine is parallelized by distributing batches of training jobs to independent worker nodes on the C3.AI suite cluster, making the physical wait time much shorter.

\subsection{Predict: Inside the \texttt{EmulatorEvaluation} pipe}\label{sec:predict}

We uniformly sample within the ranges given in Table~\ref{tab:parameters} a collection of 5,000 new parameter vectors $u^k$ (in contrast with 221 training vectors) to obtain a testing sample
\begin{equation*}
    D_{\text{test}}=\{(x, u^k, \EE[\Tilde{\eta}_{x}(u^k)\,\vert\, D_{\text{train}}], \text{Var}[\Tilde{\eta}_{x}(u^k) \,\vert\, D_{\text{train}} ]) : x\in \M, k=1,2,\ldots,5000\}.
\end{equation*}
This set pairs points in the spatiotemporal-parametric space with the emulated mean and variance of the AOD response, specifying a distribution closely mimicking the model response surface $\eta(x, u)$. We expect that this number of sampling points $u^k$ is sufficient to reliably constrain a small number of parameters. These predictions are performed efficiently by a map-reduce job across the collection of models $\Tilde{\eta}_x$.

The surrogate model appears to perform well at most locations in the spatio-temporal domain. However, gross discrepancies between the observations and the model output arise in a small fraction of the grid points which cannot be accounted for using our model discrepancy term (described later). In particular, when we consider the distance metric
\begin{equation*}
    J_{x,\delta}(u^k) = \frac{\lvert\EE[\Tilde{\eta}_{x}(u^k) \,\vert\, D_{\text{train}}] - z(x)\rvert}{\sqrt{\text{Var}[\Tilde{\eta}_{x}(u^k) \,\vert\, D_{\text{train}}] + \text{Var}[\epsilon_{\text{meas},x}] + \gamma^2}},
\end{equation*}
we identify about 2\% of the grid points for which $J_{x,\delta}(u^k)$ is a visible outlier for all $u^k$, $k=1,\ldots, 5000$. The parameter $\gamma$ here was tuned visually based on QQ plots since the other variance parameter $\delta$ is yet unestimated. Let $\M^*$ be the remaining coordinates in the spatiotemporal grid which have not been excluded either as outliers or due to missingness.

\subsection{Discrepancy: Inside the \texttt{DataDrivenModelDiscrep} pipe}\label{sec:discrep}

The quantity $\delta^2$ as seen in \eqref{eq:data_model} is the variance of the unaccounted-for uncertainty $\epsilon_{\text{other,}x}$ in our model, which we estimate using maximum likelihood. We write down the likelihood for unknowns $u$ and $\delta^2$ from our Gaussian assumption,
\begin{align*}
    L(u, \delta^2; D_{\text{train}}) = \prod_{x\in\M^*} \frac{1}{\sqrt{2\pi}} \left(\sigma^2_{x,\text{emu}}(u) + \sigma^2_{\text{meas},x} + \delta^2\right)^{-\frac12} \exp\left\{-\frac{1}{2} \frac{\left[\hat{\eta}_x(u) - z(x)\right]^2}{\sigma^2_{x,\text{emu}}(u) + \sigma^2_{\text{meas},x} + \delta^2} \right\},
\end{align*}
where
\begin{align*}
    \sigma^2_{x,\text{emu}}(u) = \text{Var}[\Tilde{\eta}_{x}(u) \,\vert\, D_{\text{train}}], \quad
    \sigma^2_{\text{meas},x} = \text{Var}[\epsilon_{\text{meas},x}], \quad
    \hat{\eta}_x(u) = \EE[\Tilde{\eta}_{x}(u)\,\vert\, D_{\text{train}}].
\end{align*}
To numerically obtain the maximum likelihood estimate for $\delta^2$, we compute the maximizing value $\hat{\delta}^2_k$ of $\log L$ for each of the test parameters $u^k$ using the scipy Python library's implementation of Brent's algorithm \citep{press_numerical_1992}; then among these maximizing values we select the one which gives the overall maximum likelihood over $k=1,2,\ldots,5000$. The resulting estimate $\hat{\delta}^2_{\text{MLE}}= \hat{\delta}^2_{\hat{k}}$, where $\hat{k} = \argmax_k \log L(u^k,\hat{\delta}^2_k; D_{\text{train}})$, is an approximate estimator, where the approximation is due to the search over the parameter space being finite.

This part of the pipeline runs quickly. Evaluating the expression for the likelihood and performing the optimization routine took about five minutes in real time in our case and can be performed in local memory. Our value for the estimator on the remaining data was $\hat{\delta}^2_{\text{MLE}} = 0.025$, which is of similar magnitude as the average measurement variance of $\Bar{\sigma}^2_{\text{meas}} = 0.027$ and emulation variance of $\Bar{\sigma}^2_{\text{emu}}(\hat{u}) = 0.038$.

\subsection{Test: Inside the \texttt{PlausibilityTest} pipe}\label{sec:test}

To obtain a confidence set for the underlying atmospheric parameters, we perform a test for parameter plausibility using MODIS AOD observations on the \texttt{MatchedGrid}. Following the terminology used in \cite{johnson_robust_2020}, we write down the \textit{implausibility metric}
\begin{equation*}
    I(u) = \sqrt{\sum_{x\in\M^*} \left(\frac{\EE[\Tilde{\eta}_{x}(u)\,\vert\, D_{\text{train}}] - z(x)}{\sqrt{\text{Var}[\Tilde{\eta}_{x}(u) \,\vert\, D_{\text{train}}] + \text{Var}[\epsilon_{\text{meas},x}] + \hat{\delta}^2_{\text{MLE}}}}\right)^2}.
\end{equation*}
For each $u$ in $D_{\text{test}}$, we compare our observed implausibility measure against its approximate distribution under the null hypothesis that $u$ is the correct parameter, $H_0: I(u) \sim \sqrt{\chi^2(df=\lvert\M^*\rvert)}.$ Here we use the facts that the sum of $n$ squared independent standard Gaussian random variables is distributed as $\chi^2(df=n)$ and that $n$ is assumed to be large enough that we can ignore the variation in the model-discrepancy variance estimate $\hat{\delta}^2_{\text{MLE}}$. In other words, testing at the 0.05 significance level, a parameter vector $u$ will be deemed implausible if $I(u)$ exceeds the $95$th percentile of the above null distribution.

A method for obtaining confidence sets inspired by the application of history matching in \cite{johnson_robust_2020} can also be derived. We find that inference based on this method is sensitive to the choice of tolerance level that is explicit in their choice of implausibility statistic. For a discussion of how our test differs from this instance of history matching, see the \hyperref[sec:appendix]{Appendix}.

\subsection{Infer: Inside the \texttt{FrequentistConfSet} pipe}\label{sec:confidence}

\begin{figure}
    \centering
    \includegraphics[width=0.9\textwidth,keepaspectratio]{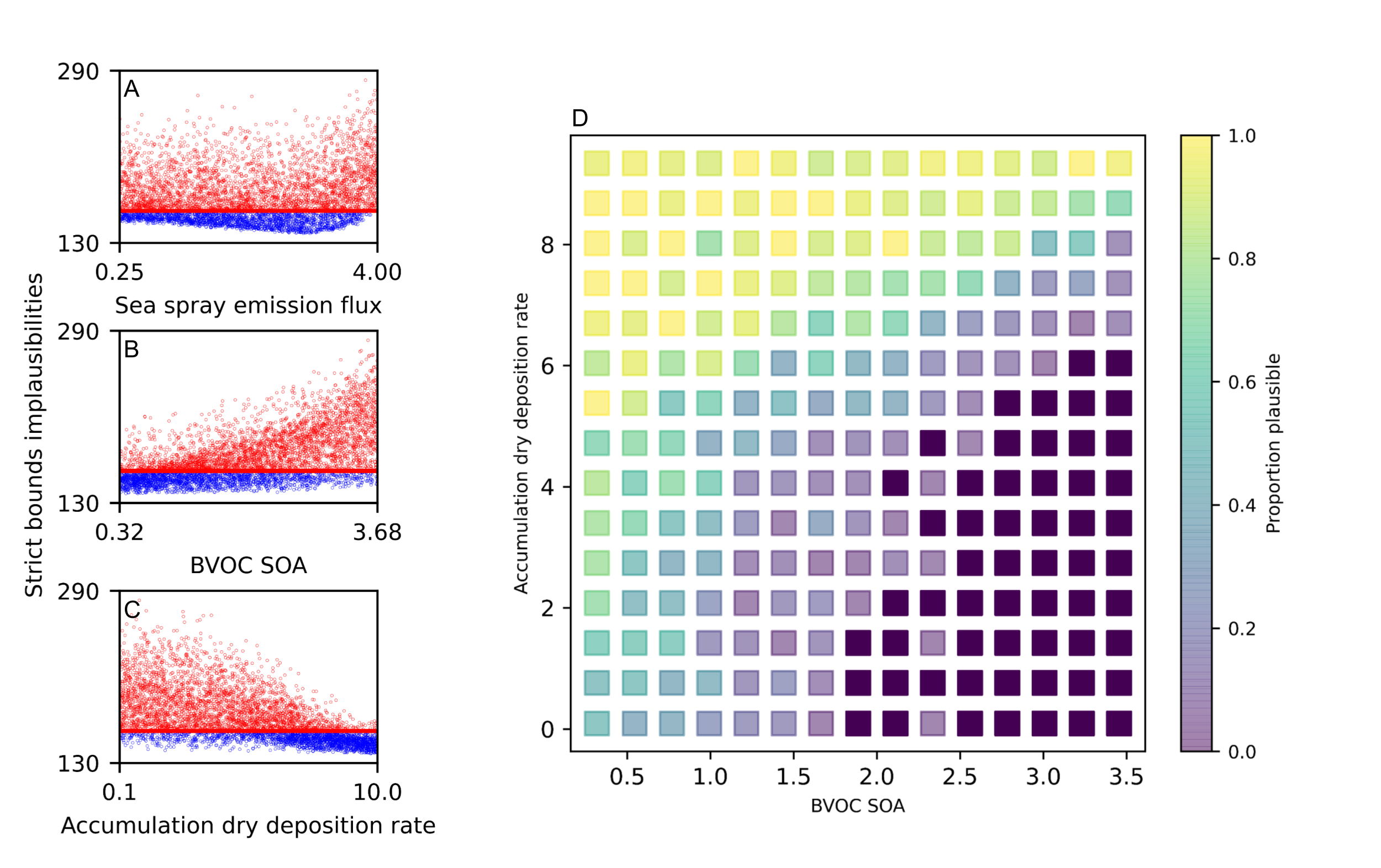}
    
    \caption{Parameter constraints at 95\% confidence level. (A-C) One-dimensional projections of the \texttt{FrequentistConfSet} described in Section~\ref{sec:confidence}. The $95$th percentile of the approximate null distribution $H_0$ is indicated by the horizontal red lines. The sea spray emission flux parameter appears to be on the verge of being constrained on its own from only two weeks of data. (D) The space spanned by the BVOC SOA and accumulation mode dry deposition rate parameters is binned, and the color of each bin shows the proportion of plausible parameter values inside. Dark purple indicates a proportion of zero---evidently, the lower right corner of this space is ruled out as implausible}
    \label{fig:triparametric_constraint}
\end{figure}

Having obtained a collection of test results for each $u^k$, $k=1,\ldots,5000$, we approximate the Neyman inversion \citep{dalmasso_likelihood-free_2023} of the test for plausibility described above by retaining all those parameters $u^k$ for which we do not reject the null at 0.05 significance level to obtain an approximate $95\%$ confidence set.\footnote{Note that the repetition of this singular hypothesis test is for inversion purposes. Since we are not inverting more than one distinct test, we do not face multiple testing issues of controlling Type I error.} This is the region of the 17-dimensional parameter space which exclusively contains non-implausible parameter vectors. A 2-dimensional projection of the resulting set is depicted on the right of Figure~\ref{fig:triparametric_constraint}.

\section{Results}\label{sec:results}

Using our strict bounds-based test for parameter plausibility, we obtain a simultaneous 95\% confidence set on the selected climate model parameters. We find that large values of the sea spray emission flux parameter are on the verge of being constrained with just two weeks of AOD data, shown in the upper left panel of Figure~\ref{fig:triparametric_constraint}.  However, the two plots on the bottom left show that for no level of either BVOC SOA or the accumulation mode dry deposition rate does our test for plausibility always fail, and so formal constraints on these cannot be obtained. To illustrate this point, suppose that the value of the accumulation dry deposition rate parameter (bottom left of Figure \ref{fig:triparametric_constraint}) was wrongly set to 1 when the true value is 10. Then there are combinations of the other 16 selected parameters that enable us to fit the model outputs to the observed AOD data within the modeled uncertainties. Hence we cannot rule out value 1 for the accumulation dry deposition rate parameter, and likewise for the other values for each parameter. If evaluated at a lower confidence level, our results seem broadly consistent with \cite{regayre_identifying_2023}, who constrain the dry deposition rate toward higher values, and \cite{regayre_value_2020} and \cite{johnson_robust_2020}, who also constrain the sea spray emissions parameter.

We are able to obtain a constraint at 95\% confidence level on the combination of the deposition rate of dry aerosols in the accumulation mode and biogenic secondary organic aerosol from volatile organic compounds. See the right panel of Figure \ref{fig:triparametric_constraint} for a binned projection of the resulting confidence set onto their span. The resulting constraint can be understood as physically meaning the following: If one hypothesizes that there are a lot of aerosols emitted by vegetation while the deposition rate is low for relatively large particles, then one will overestimate AOD so much that even when controlling for the uncertainty of the MODIS retrieval estimates due to instrumental error, the imperfection of our surrogate model, and any other sources of model discrepancy that we estimate for our climate model, a significant region of the subspace of these parameters can be ruled as implausible at the 95\% confidence~level.

\section{Conclusion}

To our knowledge, this is the first use of simulated AOD from a climate model at a time resolution as high as three-hourly to obtain observation-based constraints on input parameters likely to regulate AOD. That a salient and meaningful constraint has been gleaned from just two weeks' worth of data is suggestive of promising uses of high time resolution data in the future. Our framework is well suited for problem settings where a perturbed parameter ensemble is available for one's climate simulator and where Gaussian process emulation is appropriate. Notably, any unquantified sources of uncertainty in this setting are accounted for by the data-driven model discrepancy built into the presented pipeline, an aspect which differs from other recent scalable frameworks for model calibration, such as ESEm \citep{watson-parris_model_2021}. Our approach assumes that the model discrepancy can be captured by an additive Gaussian error that is independent across space and time so our constraints rely on these assumptions being at least approximately satisfied. A fundamental difference between our approach and the previous history matching approach as done by \cite{johnson_robust_2020} is that our method provides frequentist confidence sets with well-defined probabilistic guarantees. In addition, as described in the \hyperref[sec:appendix]{Appendix}, our method has a potential advantage over \cite{johnson_robust_2020} in that the latter is sensitive to the tuning of its tolerance level. The computational cost of our pipeline is dominated by the Gaussian process computations, so the approach is computationally feasible as long as enough parallel processing resources are available for training and evaluating the pixelwise emulators.

There are limitations to what can be achieved: with an imperfect and over-parameterized model, constraints can become inconsistent, or different parameter combinations can yield the same simulated ERF \citep{lee_relationship_2016}. However, employing more quantities for which we have observational and simulated data would nonetheless allow us to constrain a larger number of these parameters and get the most stringent constraints on aerosol radiative forcing we can. Other observable atmospheric quantities, such as sulfates or organic carbon, are sensitive to different sets of atmospheric parameters than those to which aerosol optical depth is sensitive, yielding potential opportunities to further constrain the parameter space using larger, more diverse observational data sets.

\paragraph{Acknowledgements}

We are grateful to the two anonymous referees for the insightful comments we received through the peer review process for this manuscript. We also thank the Carnegie Mellon University Statistical Methods for the Physical Sciences (STAMPS) Research Group for helpful discussions and feedback at various stages throughout this work.

\paragraph{Author Contributions}

Conceptualization: H.G., M.K.; Formal analysis: J.C.; Methodology: J.C., M.K., K.C., L.R., H.G.; Climate simulations: L.R., K.C., L.D., P.S.; Analysis software: B.A., J.C.; Writing--original draft: J.C. Writing--review and editing: J.C., M.K., H.G., L.R., P.S.

\paragraph{Competing Interests}

The authors declare that no competing interests exist.

\paragraph{Data Availability Statement}

The MODIS AOD measurements supporting the results of this work can be accessed through the MODIS web page: \url{https://modis.gsfc.nasa.gov/data/dataprod/mod04.php} (last access: 17 January 2023). Output from the A-CURE PPE is available on the CEDA archive \citep{regayre_identifying_2023}. The code to reproduce results from this paper can be found on GitHub: \url{https://github.com/c3aidti/smoke/tree/main/climateInformatics2023} (last version: 31 March 2023).

\paragraph{Funding Statement}

J.C., M.K. and H.G. acknowledge grant funding from the C3.ai Digital Transformation Institute. M.K. was also supported in part by NSF grants DMS-2053804 and PHY-2020295, and H.G. by NASA grant 80NSSC21K1344. L.R. and K.S. acknowledge funding from NERC under grants AEROS, ACID-PRUF, GASSP and A-CURE (NE/G006172/1, NE/I020059/1, NE/J024252/1 and NE/P013406/1). The funders had no role in study design, data collection and analysis, decision to publish, or preparation of the manuscript.

\newpage

\bibliography{Carzon_et_al_2023.bib}

\begin{thebibliography}{28}
\providecommand{\natexlab}[1]{#1}
\providecommand{\url}[1]{\texttt{#1}}
\expandafter\ifx\csname urlstyle\endcsname\relax
  \providecommand{\doi}[1]{doi: #1}\else
  \providecommand{\doi}{doi: \begingroup \urlstyle{rm}\Url}\fi

\bibitem[c3_()]{c3_ai_2022}
C3 enterprise ai.
\newblock URL \url{https://c3.ai/}.

\bibitem[Biegler et~al.(2010)Biegler, Biros, Ghattas, Heinkenschloss, Keyes,
  Mallick, Marzouk, Tenorio, van Bloemen~Waanders, and
  Willcox]{biegler_large-scale_2010}
Lorenz~T. Biegler, George Biros, Omar Ghattas, Matthias Heinkenschloss, David
  Keyes, Bani Mallick, Youssef Marzouk, Luis Tenorio, Bart van
  Bloemen~Waanders, and Karen Willcox, editors.
\newblock \emph{Large-scale inverse problems and quantification of
  uncertainty}.
\newblock Wiley series in computational statistics. Wiley, 2010.
\newblock ISBN 978-0-470-69743-6 978-0-470-68586-0 978-0-470-68585-3.

\bibitem[Bower et~al.(2010-12-01)Bower, Goldstein, and
  Vernon]{bower_galaxy_2010}
Richard~G. Bower, Michael Goldstein, and Ian Vernon.
\newblock Galaxy formation: a bayesian uncertainty analysis.
\newblock \emph{Bayesian Analysis}, 5\penalty0 (4):\penalty0 619--670,
  2010-12-01.
\newblock ISSN 1936-0975.
\newblock \doi{10.1214/10-BA524}.
\newblock URL
  \url{https://projecteuclid.org/journals/bayesian-analysis/volume-5/issue-4/Galaxy-formation-a-Bayesian-uncertainty-analysis/10.1214/10-BA524.full}.

\bibitem[Craig et~al.(1997)Craig, Goldstein, Seheult, and
  Smith]{craig_pressure_1997}
Peter~S. Craig, Michael Goldstein, Allan~H. Seheult, and James~A. Smith.
\newblock Pressure matching for hydrocarbon reservoirs: A case study in the use
  of bayes linear strategies for large computer experiments.
\newblock In Constantine Gatsonis, James~S. Hodges, Robert~E. Kass, Robert
  {McCulloch}, Peter Rossi, and Nozer~D. Singpurwalla, editors, \emph{Case
  Studies in Bayesian Statistics}, volume 121, pages 37--93. Springer New York,
  1997.
\newblock ISBN 978-0-387-94990-1 978-1-4612-2290-3.
\newblock \doi{10.1007/978-1-4612-2290-3_2}.
\newblock URL \url{http://link.springer.com/10.1007/978-1-4612-2290-3_2}.
\newblock Series Title: Lecture Notes in Statistics.

\bibitem[Cranmer et~al.(2020-12)Cranmer, Brehmer, and
  Louppe]{cranmer_frontier_2020}
Kyle Cranmer, Johann Brehmer, and Gilles Louppe.
\newblock The frontier of simulation-based inference.
\newblock \emph{Proceedings of the National Academy of Sciences}, 117\penalty0
  (48):\penalty0 30055--30062, 2020-12.
\newblock ISSN 0027-8424, 1091-6490.
\newblock \doi{10.1073/pnas.1912789117}.
\newblock URL \url{https://pnas.org/doi/full/10.1073/pnas.1912789117}.

\bibitem[Dalmasso et~al.(2023-01-29)Dalmasso, Masserano, Zhao, Izbicki, and
  Lee]{dalmasso_likelihood-free_2023}
Niccolò Dalmasso, Luca Masserano, David Zhao, Rafael Izbicki, and Ann~B. Lee.
\newblock Likelihood-free frequentist inference: Confidence sets with correct
  conditional coverage, 2023-01-29.
\newblock URL \url{http://arxiv.org/abs/2107.03920}.

\bibitem[Higdon et~al.(2008-06-01)Higdon, Gattiker, Williams, and
  Rightley]{higdon_computer_2008}
Dave Higdon, James Gattiker, Brian Williams, and Maria Rightley.
\newblock Computer model calibration using high-dimensional output.
\newblock \emph{Journal of the American Statistical Association}, 103\penalty0
  (482):\penalty0 570--583, 2008-06-01.
\newblock ISSN 0162-1459, 1537-274X.
\newblock \doi{10.1198/016214507000000888}.
\newblock URL
  \url{https://www.tandfonline.com/doi/full/10.1198/016214507000000888}.

\bibitem[Hubank et~al.(2020-08-06)Hubank, Platnick, King, and
  Ridgway]{hubank_modis_2020}
Paul Hubank, Steven Platnick, Michael King, and Bill Ridgway.
\newblock {MODIS} atmosphere l3 gridded product algorithm theoretical basis
  document ({ATBD}) \& users guide, 2020-08-06.
\newblock URL
  \url{https://atmosphere-imager.gsfc.nasa.gov/sites/default/files/ModAtmo/documents/L3_ATBD_C6_C61_2020_08_06.pdf}.

\bibitem[Johansen(2008)]{johansen_statistical_2008}
Kent Johansen.
\newblock \emph{Statistical Methods for History Matching}.
\newblock PhD thesis, Technical University of Denmark, 2008.

\bibitem[Johnson et~al.(2018-09-11)Johnson, Regayre, Yoshioka, Pringle, Lee,
  Sexton, Rostron, Booth, and Carslaw]{johnson_importance_2018}
Jill~S. Johnson, Leighton~A. Regayre, Masaru Yoshioka, Kirsty~J. Pringle,
  Lindsay~A. Lee, David M.~H. Sexton, John~W. Rostron, Ben B.~B. Booth, and
  Kenneth~S. Carslaw.
\newblock The importance of comprehensive parameter sampling and multiple
  observations for robust constraint of aerosol radiative forcing.
\newblock \emph{Atmospheric Chemistry and Physics}, 18\penalty0 (17):\penalty0
  13031--13053, 2018-09-11.
\newblock ISSN 1680-7324.
\newblock \doi{10.5194/acp-18-13031-2018}.
\newblock URL \url{https://acp.copernicus.org/articles/18/13031/2018/}.

\bibitem[Johnson et~al.(2020-08-13)Johnson, Regayre, Yoshioka, Pringle,
  Turnock, Browse, Sexton, Rostron, Schutgens, Partridge, Liu, Allan, Coe,
  Ding, Cohen, Atanacio, Vakkari, Asmi, and Carslaw]{johnson_robust_2020}
Jill~S. Johnson, Leighton~A. Regayre, Masaru Yoshioka, Kirsty~J. Pringle,
  Steven~T. Turnock, Jo~Browse, David M.~H. Sexton, John~W. Rostron, Nick A.~J.
  Schutgens, Daniel~G. Partridge, Dantong Liu, James~D. Allan, Hugh Coe, Aijun
  Ding, David~D. Cohen, Armand Atanacio, Ville Vakkari, Eija Asmi, and Ken~S.
  Carslaw.
\newblock Robust observational constraint of uncertain aerosol processes and
  emissions in a climate model and the effect on aerosol radiative forcing.
\newblock \emph{Atmospheric Chemistry and Physics}, 20\penalty0 (15):\penalty0
  9491--9524, 2020-08-13.
\newblock ISSN 1680-7324.
\newblock \doi{10.5194/acp-20-9491-2020}.
\newblock URL \url{https://acp.copernicus.org/articles/20/9491/2020/}.

\bibitem[Kennedy and O'Hagan(2001)]{kennedy_bayesian_2001}
Marc~C. Kennedy and Anthony O'Hagan.
\newblock Bayesian calibration of computer models.
\newblock \emph{Journal of the Royal Statistical Society: Series B (Statistical
  Methodology)}, 63\penalty0 (3):\penalty0 425--464, 2001.
\newblock ISSN 13697412.
\newblock \doi{10.1111/1467-9868.00294}.
\newblock URL
  \url{https://onlinelibrary.wiley.com/doi/10.1111/1467-9868.00294}.

\bibitem[Lee et~al.(2016-05-24)Lee, Reddington, and
  Carslaw]{lee_relationship_2016}
Lindsay~A. Lee, Carly~L. Reddington, and Kenneth~S. Carslaw.
\newblock On the relationship between aerosol model uncertainty and radiative
  forcing uncertainty.
\newblock \emph{Proceedings of the National Academy of Sciences}, 113\penalty0
  (21):\penalty0 5820--5827, 2016-05-24.
\newblock ISSN 0027-8424, 1091-6490.
\newblock \doi{10.1073/pnas.1507050113}.
\newblock URL \url{https://pnas.org/doi/full/10.1073/pnas.1507050113}.

\bibitem[{McNeall} et~al.(2016-11-24){McNeall}, Williams, Booth, Betts,
  Challenor, Wiltshire, and Sexton]{mcneall_impact_2016}
Doug {McNeall}, Jonny Williams, Ben Booth, Richard Betts, Peter Challenor, Andy
  Wiltshire, and David Sexton.
\newblock The impact of structural error on parameter constraint in a climate
  model.
\newblock \emph{Earth System Dynamics}, 7\penalty0 (4):\penalty0 917--935,
  2016-11-24.
\newblock ISSN 2190-4987.
\newblock \doi{10.5194/esd-7-917-2016}.
\newblock URL \url{https://esd.copernicus.org/articles/7/917/2016/}.

\bibitem[Patil et~al.(2022-09-30)Patil, Kuusela, and
  Hobbs]{patil_objective_2022}
Pratik Patil, Mikael Kuusela, and Jonathan Hobbs.
\newblock Objective frequentist uncertainty quantification for atmospheric
  {\textbackslash}({\textbackslash}mathrm\{{CO}\}\_2{\textbackslash})
  retrievals.
\newblock \emph{{SIAM}/{ASA} Journal on Uncertainty Quantification},
  10\penalty0 (3):\penalty0 827--859, 2022-09-30.
\newblock ISSN 2166-2525.
\newblock \doi{10.1137/20M1356403}.
\newblock URL \url{https://epubs.siam.org/doi/10.1137/20M1356403}.

\bibitem[Press(1992)]{press_numerical_1992}
William~H. Press, editor.
\newblock \emph{Numerical recipes in C: the art of scientific computing}.
\newblock Cambridge University Press, 2nd ed edition, 1992.
\newblock ISBN 978-0-521-43108-8 978-0-521-43720-2.

\bibitem[Rasmussen and Williams(2006)]{rasmussen_gaussian_2006}
Carl~Edward Rasmussen and Christopher K.~I. Williams.
\newblock \emph{Gaussian processes for machine learning}.
\newblock Adaptive computation and machine learning. {MIT} Press, 2006.
\newblock ISBN 978-0-262-18253-9.
\newblock {OCLC}: ocm61285753.

\bibitem[Regayre et~al.(2018-07-13)Regayre, Johnson, Yoshioka, Pringle, Sexton,
  Booth, Lee, Bellouin, and Carslaw]{regayre_aerosol_2018}
Leighton~A. Regayre, Jill~S. Johnson, Masaru Yoshioka, Kirsty~J. Pringle, David
  M.~H. Sexton, Ben B.~B. Booth, Lindsay~A. Lee, Nicolas Bellouin, and
  Kenneth~S. Carslaw.
\newblock Aerosol and physical atmosphere model parameters are both important
  sources of uncertainty in aerosol {ERF}.
\newblock \emph{Atmospheric Chemistry and Physics}, 18\penalty0 (13):\penalty0
  9975--10006, 2018-07-13.
\newblock ISSN 1680-7324.
\newblock \doi{10.5194/acp-18-9975-2018}.
\newblock URL \url{https://acp.copernicus.org/articles/18/9975/2018/}.

\bibitem[Regayre et~al.(2020-08-28)Regayre, Schmale, Johnson, Tatzelt,
  Baccarini, Henning, Yoshioka, Stratmann, Gysel-Beer, Grosvenor, and
  Carslaw]{regayre_value_2020}
Leighton~A. Regayre, Julia Schmale, Jill~S. Johnson, Christian Tatzelt, Andrea
  Baccarini, Silvia Henning, Masaru Yoshioka, Frank Stratmann, Martin
  Gysel-Beer, Daniel~P. Grosvenor, and Ken~S. Carslaw.
\newblock The value of remote marine aerosol measurements for constraining
  radiative forcing uncertainty.
\newblock \emph{Atmospheric Chemistry and Physics}, 20\penalty0 (16):\penalty0
  10063--10072, 2020-08-28.
\newblock ISSN 1680-7324.
\newblock \doi{10.5194/acp-20-10063-2020}.
\newblock URL \url{https://acp.copernicus.org/articles/20/10063/2020/}.

\bibitem[Regayre et~al.(2023-02-16)Regayre, Deaconu, Grosvenor, Sexton,
  Symonds, Langton, Watson-Paris, Mulcahy, Pringle, Richardson, Johnson,
  Rostron, Gordon, Lister, Stier, and Carslaw]{regayre_identifying_2023}
Leighton~A. Regayre, Lucia Deaconu, Daniel~P. Grosvenor, David M.~H. Sexton,
  Christopher Symonds, Tom Langton, Duncan Watson-Paris, Jane~P. Mulcahy,
  Kirsty~J. Pringle, Mark Richardson, Jill~S. Johnson, John~W. Rostron, Hamish
  Gordon, Grenville Lister, Philip Stier, and Ken~S. Carslaw.
\newblock Identifying climate model structural inconsistencies allows for tight
  constraint of aerosol radiative forcing.
\newblock 2023-02-16.
\newblock \doi{10.5194/egusphere-2023-77}.
\newblock URL
  \url{https://egusphere.copernicus.org/preprints/2023/egusphere-2023-77/}.

\bibitem[Schafer and Stark(2009-09)]{schafer_constructing_2009}
Chad~M. Schafer and Philip~B. Stark.
\newblock Constructing confidence regions of optimal expected size.
\newblock \emph{Journal of the American Statistical Association}, 104\penalty0
  (487):\penalty0 1080--1089, 2009-09.
\newblock ISSN 0162-1459, 1537-274X.
\newblock \doi{10.1198/jasa.2009.tm07420}.
\newblock URL
  \url{http://www.tandfonline.com/doi/abs/10.1198/jasa.2009.tm07420}.

\bibitem[Schutgens et~al.(2017)Schutgens, Tsyro, Gryspeerdt, Goto, Weigum,
  Schulz, and Stier]{schutgens_spatio-temporal_2017}
Nick Schutgens, Svetlana Tsyro, Edward Gryspeerdt, Daisuke Goto, Natalie
  Weigum, Michael Schulz, and Philip Stier.
\newblock On the spatio-temporal representativeness of observations.
\newblock \emph{Atmospheric Chemistry and Physics}, 17\penalty0 (16):\penalty0
  9761--9780, 2017.

\bibitem[Sellar et~al.(2019-12)Sellar, Jones, Mulcahy, Tang, Yool, Wiltshire,
  O'Connor, Stringer, Hill, Palmieri, Woodward, Mora, Kuhlbrodt, Rumbold,
  Kelley, Ellis, Johnson, Walton, Abraham, Andrews, Andrews, Archibald,
  Berthou, Burke, Blockley, Carslaw, Dalvi, Edwards, Folberth, Gedney,
  Griffiths, Harper, Hendry, Hewitt, Johnson, Jones, Jones, Keeble, Liddicoat,
  Morgenstern, Parker, Predoi, Robertson, Siahaan, Smith, Swaminathan,
  Woodhouse, Zeng, and Zerroukat]{sellar_ukesm1_2019}
Alistair~A. Sellar, Colin~G. Jones, Jane~P. Mulcahy, Yongming Tang, Andrew
  Yool, Andy Wiltshire, Fiona~M. O'Connor, Marc Stringer, Richard Hill, Julien
  Palmieri, Stephanie Woodward, Lee Mora, Till Kuhlbrodt, Steven~T. Rumbold,
  Douglas~I. Kelley, Rich Ellis, Colin~E. Johnson, Jeremy Walton, Nathan~Luke
  Abraham, Martin~B. Andrews, Timothy Andrews, Alex~T. Archibald, Ségolène
  Berthou, Eleanor Burke, Ed~Blockley, Ken Carslaw, Mohit Dalvi, John Edwards,
  Gerd~A. Folberth, Nicola Gedney, Paul~T. Griffiths, Anna~B. Harper, Maggie~A.
  Hendry, Alan~J. Hewitt, Ben Johnson, Andy Jones, Chris~D. Jones, James
  Keeble, Spencer Liddicoat, Olaf Morgenstern, Robert~J. Parker, Valeriu
  Predoi, Eddy Robertson, Antony Siahaan, Robin~S. Smith, Ranjini Swaminathan,
  Matthew~T. Woodhouse, Guang Zeng, and Mohamed Zerroukat.
\newblock {UKESM}1: Description and evaluation of the u.k. earth system model.
\newblock \emph{Journal of Advances in Modeling Earth Systems}, 11\penalty0
  (12):\penalty0 4513--4558, 2019-12.
\newblock ISSN 1942-2466, 1942-2466.
\newblock \doi{10.1029/2019MS001739}.
\newblock URL
  \url{https://onlinelibrary.wiley.com/doi/abs/10.1029/2019MS001739}.

\bibitem[Stanley et~al.(2022-10-01)Stanley, Patil, and
  Kuusela]{stanley_uncertainty_2022}
Michael Stanley, Pratik Patil, and Mikael Kuusela.
\newblock Uncertainty quantification for wide-bin unfolding: one-at-a-time
  strict bounds and prior-optimized confidence intervals.
\newblock \emph{Journal of Instrumentation}, 17\penalty0 (10):\penalty0 P10013,
  2022-10-01.
\newblock ISSN 1748-0221.
\newblock \doi{10.1088/1748-0221/17/10/P10013}.
\newblock URL
  \url{https://iopscience.iop.org/article/10.1088/1748-0221/17/10/P10013}.

\bibitem[Stark(1992)]{stark_inference_1992}
Philip~B. Stark.
\newblock Inference in infinite-dimensional inverse problems: Discretization
  and duality.
\newblock \emph{Journal of Geophysical Research}, 97:\penalty0 14055, 1992.
\newblock ISSN 0148-0227.
\newblock \doi{10.1029/92JB00739}.
\newblock URL \url{http://doi.wiley.com/10.1029/92JB00739}.

\bibitem[Telford et~al.(2008)Telford, Braesicke, Morgenstern, and
  Pyle]{telford_technical_2008}
P~J Telford, P~Braesicke, O~Morgenstern, and J~A Pyle.
\newblock Technical note: Description and assessment of a nudged version of the
  new dynamics uniﬁed model.
\newblock \emph{Atmospheric Chemistry and Physics}, 8\penalty0 (6):\penalty0
  1701--1712, 2008.

\bibitem[Verly and Division(1984)]{verly_geostatistics_1984}
G.~Verly and North Atlantic Treaty Organization Scientific~Affairs Division.
\newblock \emph{Geostatistics for Natural Resources Characterization}.
\newblock Number pt. 2 in Geostatistics for Natural Resources Characterization.
  D. Reidel Publishing Company, 1984.
\newblock ISBN 978-90-277-1747-4.
\newblock URL \url{https://books.google.com/books?id=x2gZAQAAIAAJ}.

\bibitem[Watson-Parris et~al.(2021-12-20)Watson-Parris, Williams, Deaconu, and
  Stier]{watson-parris_model_2021}
Duncan Watson-Parris, Andrew Williams, Lucia Deaconu, and Philip Stier.
\newblock Model calibration using {ESEm} v1.1.0 – an open, scalable earth
  system emulator.
\newblock \emph{Geoscientific Model Development}, 14\penalty0 (12):\penalty0
  7659--7672, 2021-12-20.
\newblock ISSN 1991-9603.
\newblock \doi{10.5194/gmd-14-7659-2021}.
\newblock URL \url{https://gmd.copernicus.org/articles/14/7659/2021/}.

\end{thebibliography}

\newpage

\section*{Appendix: Comparison with the method of history matching}\label{sec:appendix}

It is possible to provide frequentist confidence sets on parameters using a method inspired by history matching. Below, we describe an adaptation of the version of the method employed in \cite{johnson_robust_2020}. Principally, the typical use of history matching in that work and others is to exclude implausible parameter values in a systematic way, not to produce a plausible constraint set with fixed probabilistic properties.

One difference between the method of \cite{johnson_robust_2020} and our \textsc{PlausibilityTest} pipe of Section~\ref{sec:test} is the choice of implausibility statistic. \cite{johnson_robust_2020} use the statistic
\begin{equation*}
    I_{\text{HM}, N}(u) = \left\{\frac{\lvert\EE[\Tilde{\eta}_{x}(u)\,\vert\, D_{\text{train}}] - z(x)\rvert}{\sqrt{\text{Var}[\Tilde{\eta}_{x}(u) \,\vert\, D_{\text{train}}] + \text{Var}[\epsilon_{\text{meas},x}] + \text{Var}[\epsilon_{\text{other},x}]}} : x\in\M^*\right\}_{(N)},
\end{equation*}
where $S_{(N)}$ denotes the $N$th largest element of a set $S$. For instance,
\begin{equation*}
    I_{\text{HM}, 1}(u) = \max \left\{\frac{\lvert\EE[\Tilde{\eta}_{x}(u)\,\vert\, D_{\text{train}}] - z(x)\rvert}{\sqrt{\text{Var}[\Tilde{\eta}_{x}(u) \,\vert\, D_{\text{train}}] + \text{Var}[\epsilon_{\text{meas},x}] + \text{Var}[\epsilon_{\text{other},x}]}} : x\in\M^*\right\}.
\end{equation*}
The authors then tune the ``tolerance level'' (which corresponds to the number $(\lvert \M^*\rvert - N)$ in our expression for the statistic) and the ``exceedence threshold'' (which, by analogy to our test, corresponds to an implausibility cutoff or critical value; in most cases set to $3.5$) until a fixed proportion (say, $40\%$) of test parameters $u$ are retained as plausible under the statistic. By this account, a constraint on the parameters is necessarily yielded, but the confidence level at which this constraint holds is undetermined.

\begin{figure}
    \centering
    \includegraphics[width=0.9\textwidth,keepaspectratio]{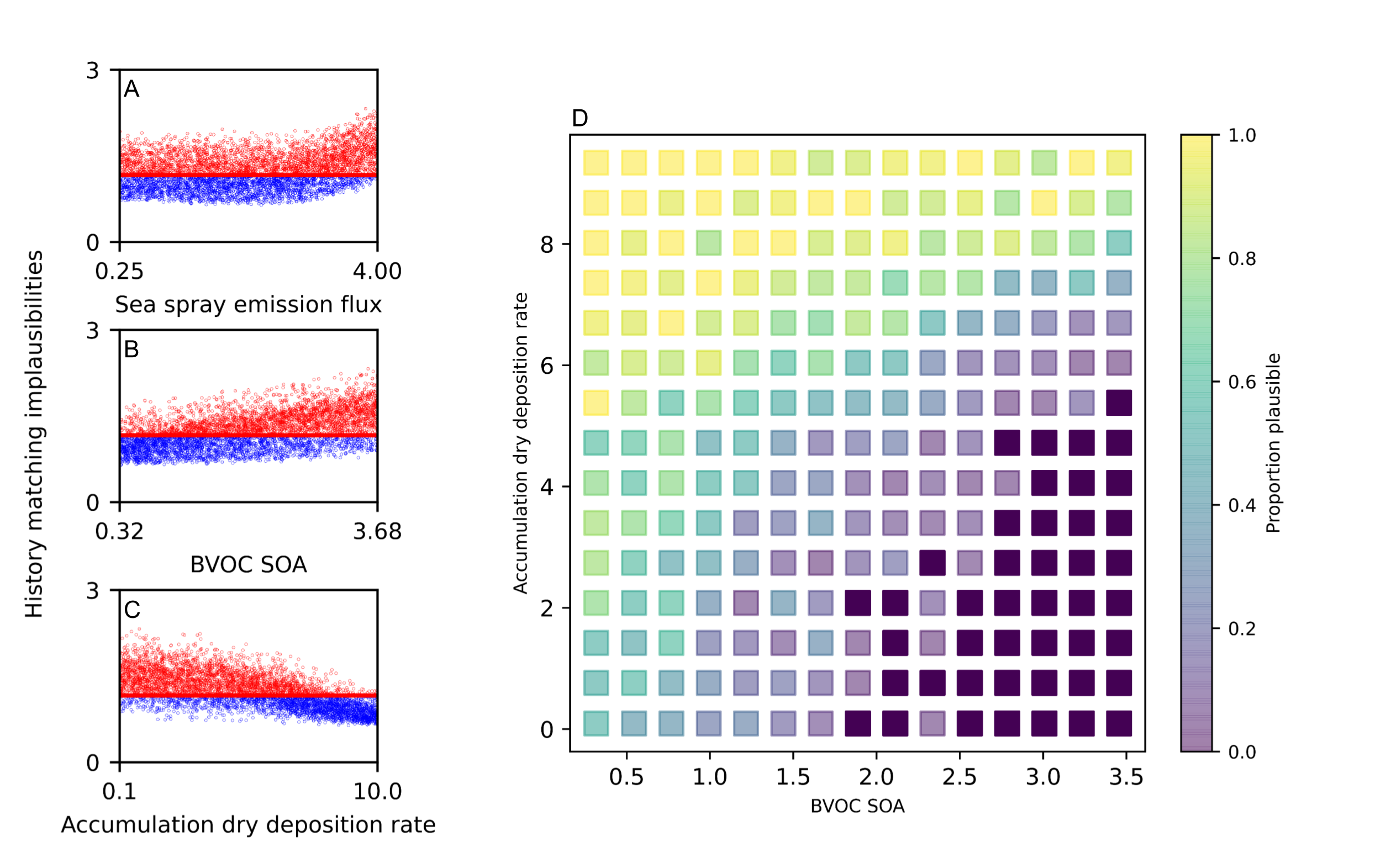}
    \caption{Constraints resulting from an adaptation of the history matching method at $95\%$ confidence level. Referring to the implausibility statistic given by Eq.~\eqref{eq:hm_implausibility_metric}, we use $q=0.25$}
    \label{fig:history_matching}
\end{figure}

Alternatively, we can set a confidence level first and compare the resulting constraints obtained by the strict bounds approach or this history matching-inspired approach, where the former approach yields the constraints shown in Figure~\ref{fig:triparametric_constraint}. In particular, consider the implausibility statistic
\begin{equation}\label{eq:hm_implausibility_metric}
    I_{1-q}(u) = \text{quantile}_{1-q} \left\{\frac{\lvert\EE[\Tilde{\eta}_{x}(u)\,\vert\, D_{\text{train}}] - z(x)\rvert}{\sqrt{\text{Var}[\Tilde{\eta}_{x}(u) \,\vert\, D_{\text{train}}] + \text{Var}[\epsilon_{\text{meas},x}] + \hat{\delta}^2_{\text{MLE}}}} : x\in\M^* \right\},
\end{equation}
where $\text{quantile}_{1-q}$ returns the $(1-q)$th quantile of the given set. For example, $I_1(u)$ returns the maximum absolute normalized discrepancy for parameter $u$ and $I_{0.5}(u)$ is the median. This is a close analog to the statistic seen in \cite{johnson_robust_2020} when $q\approx N/\lvert \M^*\rvert$. Naturally the approximate null distribution for this new statistic is that of the $(1-q)$th percentile of a sample of $\lvert\M^*\rvert$ half-normal random variables. A critical value for the plausibility test at the $5\%$ significance level can therefore be estimated quickly by simulating a collection of samples of $\lvert\M^*\rvert$ half-normal random variables, drawing the $(1-q)$th percentile from each sample, and then selecting the $95$th percentile from that collection. In Figure \ref{fig:history_matching}, we show the $95\%$ confidence level constraints on the aerosol parameters that are obtained from the above-described history matching method using parameter $q=0.25$.

As this example illustrates, similar non-trivial and principled constraints on aerosol parameters are possible. However, the history matching-inspired approach requires choosing the tuning parameter $q$ which substantially affects the final constraints. In Figure~\ref{fig:history_matching}, we chose $q$ to obtain constraints similar to those given by our method in Figure~\ref{fig:triparametric_constraint}. If we choose $q = 0.5$ (i.e., use the median), we find that the constraints (not shown) become looser than those provided by our method. On the other hand, if we choose $q$ close to zero, we find that this history matching approach becomes sensitive to non-Gaussianities in the tails of the error distributions, leading to overly discriminating plausiblity test results. If history matching was calibrated like this to obtain confidence sets at a prescribed confidence level (which, we emphasize, is not currently done), it seems difficult to choose $q$ optimally to balance the power of the tests with robustness to mismodeling of the error distribution tails.

\end{document}